\documentclass[onecolumn,showpacs,preprintnumbers,amsmath,amssymb]{revtex4}
\usepackage{graphicx}
\usepackage{dcolumn}
\usepackage{bm}
\begin{document}


\title{Magnetic Fields of Spherical Compact Stars in Braneworld}

\author{B.~J.~Ahmedov}
\email{ahmedov@astrin.uzsci.net}
\author{ F.~J.~Fattoyev}%
 \email{farid@ictp.it}
\affiliation{%
Institute of Nuclear Physics and
    Ulugh Beg Astronomical Institute, Astronomicheskaya 33,
    Tashkent 700052, Uzbekistan \\
    The Abdus Salam International Centre
for Theoretical Physics, 34014 Trieste, Italy}%

\date{\today}

\begin{abstract}
\noindent \noindent We study the dipolar magnetic field
configuration in dependence on brane tension and present solutions
of Maxwell equations in the internal and external background
spacetime of a magnetized spherical star in a Randall-Sundrum II
type braneworld. The star is modelled as sphere consisting of
perfect highly magnetized fluid with infinite conductivity and
frozen-in dipolar magnetic field. With respect to solutions for
magnetic fields found in the Schwarzschild spacetime brane tension
introduces enhancing corrections both to the interior and the
exterior magnetic field. These corrections could be relevant for
the magnetic fields of magnetized compact objects as pulsars and
magnetars and may provide the observational evidence for the brane
tension through the modification of formula for magneto-dipolar
emission which gives amplification of electromagnetic energy loss
up to few orders depending on the value of the brane tension.
\end{abstract}

\maketitle

\section{Introduction}

It is well known that magnetic fields play an important role in
the life history of majority astrophysical objects especially of
compact relativistic stars which posses surface magnetic fields of
$10^{12}G$. Magnetic fields of magnetars~\cite{dt92,td95} can
reach up to $10^{15}\div 10^{16}G$ and in the deep interior of
compact stars magnetic field strength may be estimated up to
$10^{18}G$. The strength of compact star's magnetic field is one
of the main quantities determining their observability, for
example as pulsars through magneto-dipolar radiation. Therefore it
is extremely important to study the effect of the different
phenomena on evolution and behaviour of stellar interior and
exterior magnetic fields.

Recently obtained solutions for relativistic stars on
branes~\cite{mm05}--\cite{nkd00} have created an interest in the
study of the influence and implications of the bulk tension on the
various astrophysical processes, for example to the gravitational
lensing~\cite{k06,kp05a,kp05b} and to the motion of test
particles~\cite{i05}. To the best of our knowledge the effect of
brane tension on magnetic field configuration in compact stars has
not yet been studied. Since magnetic field determines the reach
observational phenomenology of compact stars we will study here
the consequences of the brane effects on stellar magnetic fields.

In the Newtonian framework the exterior electromagnetic fields of
magnetized and rotating sphere are given in the classical paper of
Deutsch~\cite{d55} and interior fields are studied by many
authors, for example, in the paper~\cite{rt73}. In the
general-relativistic approach the study of the  magnetic field
structure outside magnetized  compact gravitational objects has
been initiated by the pioneering work of Ginzburg and
Ozernoy~\cite{go64} and have been further extended by number of
authors~\cite{ac70}--\cite{japan}, while in some
papers~\cite{getal98}--\cite{zr} the work has been completed by
considering magnetic fields interior relativistic star for the
different models of stellar matter. General-relativistic treatment
for the structure of external and internal stellar magnetic fields
including numerical results has shown that the magnetic field is
amplified by the monopolar part of gravitational field depending
on the compactness of the relativistic star.

Here we will consider static spherically symmetric stars in the
braneworld endowed with strong magnetic fields. The magnetic field
structure will be assumed to be dipolar and axisymmetric and the
effect of the gravitational field of the star in the braneworld on
the magnetic field structure is considered without feedback,
amounting to the astrophysical evidence that the magnetic field
energy is not strong enough to affect the spacetime geometry. We
find an exact analytical interior solution to Maxwell equations
for the magnetic field when the stellar matter has unrealistic
equation of state of stiff matter. When stellar matter has
constant density the interior fields are found as numerical
solutions of Maxwell equations. External magnetic fields are also
found numerically. We show that both external and interior
magnetic fields will be essentially modified by five-dimensional
gravity effects.

In the paper we describe our model assumptions, present the set of
the Maxwell equations to be solved together with boundary and
initial conditions for the models under consideration and obtain
analytical and numerical results for the stellar magnetic field.
The paper is organized as follows. In section \ref{meq} we provide
a description of spherical compact objects in the braneworld and
basic Maxwell's equations in the spacetime of these objects. In
section \ref{ss} we make main assumptions, discuss the boundary
and initial conditions for the magnetic fields and also discuss
interior magnetic fields for the different equations of state. In
subsection~\ref{internan} we find exact analytical solution for
the stellar matter with the stiff equation of state. In
subsection~\ref{srst_es} we integrate external Maxwell equations
from asymptotical infinity to the surface of star and find
numerical solutions for magnetic field outside the braneworld
star. In subsection~\ref{internum} we numerically integrate the
equation for magnetic field from the surface of the star till some
inner radius.  As astrophysical application of the obtained
results we look for the modification of luminosity of
electromagnetic magneto-dipolar radiation from the rotating star
due the braneworld effects in section~\ref{application}.

Throughout, we use a space-like signature $(-,+,+,+)$ and a system
of units in which $G = 1 = c$ (However, for those expressions with
an astrophysical application we have written the speed of light
explicitly.). Greek indices are taken to run from 0 to 3 and Latin
indices from 1 to 3; covariant derivatives are denoted with a
semi-colon and partial derivatives with a comma.

\section{Maxwell Equations In a Spacetime of Spherical Star
in the Braneworld}
\label{meq}

The braneworld paradigm views our universe as a slice of some
higher dimensional spacetime, in which we have standard four
dimensional physics confined to the brane, and only gravitational
plus few number of other fields can propagate in the bulk. The
difficulty in analysis of gravitational field equations and
gravitational collapse on the branes is arising from the fact that
the propagation of gravity into the bulk does not provide an
opportunity to have brane equations for gravitational field in
closed form~\cite{mm05}.

In a coordinate system $(ct,r,\theta,\phi)$, the space-time metric
for a spherical relativistic star in the braneworld is
{\cite{mm05}--\cite{gm01}}
\begin{equation}
\label{slow_rot}
ds^2 = -A^2(r) dt^2 + H^2(r)dr^2 +
    r^2 d\theta ^2+ r^2\sin^2\theta d\phi ^2 \ .
\end{equation}
and Einstein equations on brane for unknown metric functions
$A(r)$ and $H(r)$ imply
\begin{eqnarray}
\label{eins1}&&
\frac{1}{r^2}-\frac{1}{H^2}\left(\frac{1}{r^2}-\frac{2}{r}\frac{H'}{H}\right)
= 8 \pi  \rho^{eff} \ , \\ \
\label{eins2} &&
-\frac{1}{r^2}+\frac{1}{H^2}\left(\frac{1}{r^2}+\frac{2}{r}\frac{A'}{A}\right)
= 8 \pi  \left(\rho^{eff}+\frac{4}{k^2 \lambda}P\right) \ , \\ \
\label{eins3}
&& p'+\frac{A'}{A}(\rho+p)=0 \ , \\ \
\label{eins4}&&
U'+ 4\frac{A'}{A}U+2P'
+2\frac{A'}{A}P+\frac{6}{r}P= -2(4\pi)^2(\rho+p)\rho' \ .
\end{eqnarray}
The nonlocal bulk effects are carried by the nonlocal energy
density $U$ and nonlocal pressure $P$. Here prime denotes the
radial derivative, $\rho^{eff}$ is the effective total energy
density, $\rho$ and $p$ are matter energy density and pressure,
respectively.

Outside the star, an exact solution for the metric
(\ref{slow_rot}) is completely known and explicit expressions for
the metric functions are given by the Reissner-Nordstr\"{o}m-type
solution found in \cite{nkd00}
\begin{equation}
A^2(r) \equiv
    \left(1-\frac{2 M}{r}+\frac{Q}{r^2}\right)
    = H^{-2}(r) \ , \hskip 1.0cm
    r > R \ ,
\end{equation}
in which there is no electric charge, but instead a negative Weyl
``charge'' $Q=-3MR\rho/\lambda$ is arising from bulk gravitational
tidal effects as the projection of gravitational field on to the
brane. Here $M$ and $R$ are the total mass and radius of the star,
$\lambda$ is the brane tension. Classical general relativity is
regained when $\lambda^{-1}\rightarrow 0$.

    The bulk tidal charge strengthens the gravitational field of
the relativistic stars and black holes. The metric with the
negative tidal charge $Q$ has a space-like singularity and horizon
is given by the radius
\begin{equation}
r_+ = M \left(1+\sqrt{1-\frac{Q}{M^{2}}}\right)
\end{equation}
which is larger than the Schwarzschild horizon.

    The general form of the first pair of general
relativistic Maxwell equations is given by
\begin{equation}
\label{maxwell_firstpair}
3! F_{[\alpha \beta, \gamma]} =  2 \left(F_{\alpha \beta, \gamma }
    + F_{\gamma \alpha, \beta} + F_{\beta \gamma,\alpha}
    \right) = 0 \ ,
\end{equation}
where $F_{\alpha \beta}$ is the electromagnetic field
tensor expressing the strict connection between the
electric and magnetic four-vector fields $E^{\alpha},\
B^{\alpha}$. For an observer with four-velocity
$u^{\alpha}$, the covariant components of the
electromagnetic tensor are given by
\begin{equation}
\label{fab_def}
F_{\alpha\beta} \equiv 2 u_{[\alpha} E_{\beta]} +
    \eta_{\alpha\beta\gamma\delta}u^\gamma B^\delta \ ,
\end{equation}
where $T_{[\alpha \beta]} \equiv \frac{1}{2}(T_{\alpha
\beta} - T_{\beta \alpha})$ and
$\eta_{\alpha\beta\gamma\delta}$ is the pseudo-tensorial
expression for the Levi-Civita symbol $\epsilon_{\alpha
\beta \gamma \delta}$
\begin{equation}
\eta^{\alpha\beta\gamma\delta}=-\frac{1}{\sqrt{-g}}
    \epsilon_{\alpha\beta\gamma\delta} \ ,
    \hskip 2.0cm
\eta_{\alpha\beta\gamma\delta}=
    \sqrt{-g}\epsilon_{\alpha\beta\gamma\delta} \ ,
\end{equation}
with $g\equiv {\rm det}|g_{\alpha\beta}|=-A^2H^2 r^4 \sin^2\theta$
for the metric (\ref{slow_rot}).

For the stationary proper observers four-velocity components given
by
\begin{equation}
\label{uzamos}
(u^{\alpha})_{_{\rm obs}}\equiv
    A^{-1}\bigg(1,0,0,0\bigg) \ ;
    \hskip 2.0cm
(u_{\alpha})_{_{\rm obs}}\equiv
    A \bigg(- 1,0,0,0 \bigg) \ .
\end{equation}

    In the spacetime (\ref{slow_rot}) and
with the definition (\ref{fab_def}) referred to the
observers (\ref{uzamos}), the first pair of Maxwell
equations (\ref{maxwell_firstpair}) take then the form
\begin{eqnarray}
\label{max1_ea}
&& \left(Hr^2\sin\theta B^i \right)_{,i}=0 \ ,
\\
\label{max1_eb}
&& \left(Hr^2\sin\theta \right)\frac{\partial B^{r}}{\partial t}
    = A\left(E_{\theta,\phi}- E_{\phi ,\theta} \right) \ ,
\\
\label{max1_ec}
&& \left(Hr^2\sin\theta \right)
    \frac{\partial B^{\theta}}{\partial t}
    = \left( E_{\phi} \  A\right)_{,r}-  AE_{r,\phi} \ ,
\\
\label{max1_ed}
&& \left(Hr^2\sin\theta \right)\frac{\partial B^{\phi}}{\partial t}
    = -\left( E_{\theta} \ {A}\right)_{,r}+
    {A}E_{r,\theta}\ .
\end{eqnarray}

    The general form of the second pair of Maxwell
equations is given by
\begin{equation}
\label{maxwell_secondpair}
F^{\alpha \beta}_{\ \ \ \ ;\beta} = 4\pi J^{\alpha}
\end{equation}
where the four-current $J^{\alpha}$ is a sum of
convection and conduction currents
\begin{equation}
J^{\alpha}=\rho_e w^\alpha + j^\alpha \ ,
    \hskip 1.0cm j^\alpha w_\alpha \equiv 0\ ,
    \qquad
    j_\alpha = \sigma F_{ \alpha \beta}w^\beta \ ,
\end{equation}
where $\sigma$ is the electrical conductivity and $w^\alpha$ is
four-velocity of conducting medium as a whole.

We can now rewrite the second pair of Maxwell
equations as
\begin{eqnarray}
\label{max2_ea}
\left({Hr^2\sin\theta} E^i\right)_{,i}
    &=& 4\pi AH r^2\sin\theta J^0 \ ,
\\
\label{max2_eb}
A \left( B_{\phi\;,\theta} -  B_{\theta\;,\phi} \right)
    &=& \left(Hr^2\sin\theta \right)
    \frac{\partial E^r}{\partial t}
    +4\pi AH r^2\sin\theta J^r\ ,
\\
\label{max2_ec}
{A}B_{r,\phi}-\left({A} B_\phi \right)_{,r}
    &=& \left(Hr^2\sin\theta
    \right)\frac{\partial E^\theta}{\partial t}
    +4\pi AH r^2\sin\theta J^\theta \ ,
\\
\label{max2_ed}
\left({A} B_\theta \right)_{,r} - AB_{r\;,\theta}
    &=& \left(Hr^2\sin\theta \right)
    \frac{\partial E^\phi}{\partial t} +
    4\pi AH r^2\sin\theta J^\phi \ .
    \hskip 1.5 cm
\end{eqnarray}

In order to obtain physical components of quantities, one has to
select a tetrad frame to which the components of the electromagnetic
field to be projected. The orthonormal tetrad is a set of four
mutually orthogonal unit vectors at each point in a given spaciteme
which gives the directions of the four axes of locally Minkowskian
coordinate system. Such an orthonormal tetrad associated with the
metric~(\ref{slow_rot})
$\{{\mathbf e}_{\hat \mu}\} = ({\mathbf e}_{\hat 0}, {\mathbf
e}_{\hat r}, {\mathbf e}_{\hat \theta}, {\mathbf e}_{\hat \phi})$
and carried by a proper observer (\ref{uzamos}) can be selected as
\begin{eqnarray}
\label{zamo_tetrad_0}
{\mathbf e}_{\hat 0}^{\alpha}  =
    A^{-1}\bigg(1,0,0,0\bigg) \ ,   \quad
{\mathbf e}_{\hat r}^{\alpha}  =
    H^{-1}\bigg(0,1,0,0\bigg) \ ,       \quad
{\mathbf e}_{\hat \theta}^{\alpha}  =
    \frac{1}{r}\bigg(0,0,1,0\bigg)  \ ,         \quad
\nonumber \\ 
{\mathbf e}_{\hat \phi}^{\alpha}  =
    \frac{1}{r\sin\theta}\bigg(0,0,0,1\bigg) \ .
\end{eqnarray}

\noindent The 1-forms $\{{\mathbf \omega}^{\hat
\mu}\} = ({\mathbf \omega}^{\hat 0}, {\mathbf
\omega}^{\hat r}, {\mathbf \omega}^{\hat \theta},
{\mathbf \omega}^{\hat \phi})$, corresponding to this
tetrad have instead components
\begin{eqnarray}
\label{zamo_tetrad_1-forms_0}
\mathbf{\omega}^{\hat 0}_{\alpha}  =
    A\bigg(1,0,0,0\bigg)\ ,         \quad
\mathbf{\omega}^{\hat r}_{\alpha}  =
    H\bigg(0,1,0,0\bigg)\ ,         \quad
\mathbf{\omega}^{\hat \theta}_{\alpha}  =
    {r}\bigg(0,0,1,0\bigg)\ ,           \quad
\nonumber \\ 
\mathbf{\omega}^{\hat \phi}_{\alpha}  =
    {r\sin\theta}\bigg(0,0,0,1\bigg)\ .
\end{eqnarray}

    We can now rewrite Maxwell equations
(\ref{max1_ea})--(\ref{max1_ed}) and
(\ref{max2_ea})--(\ref{max2_ed}) in the orthonormal reference
frame by contracting (\ref{maxwell_firstpair}) and
(\ref{maxwell_secondpair}) with
(\ref{zamo_tetrad_0})
and
(\ref{zamo_tetrad_1-forms_0}):
%
\begin{eqnarray}
\label{max1a}
&&\sin\theta \left(r^2B^{\hat r}\right)_{,r}+
    Hr\left(\sin\theta B^{\hat \theta}\right)_{,\theta} +
    H r B^{\hat \phi}_{\ , \phi} = 0 \ ,
\\
\label{max1b}
&&\left({r\sin\theta}\right)\frac{\partial B^{\hat r}}{\partial t}
     =  {A} \left[E^{\hat\theta}_{\ ,\phi}- \left(\sin\theta
    E^{\hat \phi} \right)_{,\theta}\right]\ ,
\\
\label{max1c}
&&\left({Hr\sin\theta}\right)
    \frac{\partial B^{\hat \theta}}{\partial t}
    = -AH E^{\hat r}_{\ ,\phi} +
    \sin\theta \left(r A E^{\hat \phi} \right)_{,r}\ ,
\\
\label{max1d}
&&\left({Hr}\right)
    \frac{\partial B^{\hat \phi}}{\partial t}
    = - \left(r A E^{\hat \theta}\right)_{,r}
    + AH E^{\hat r}_{ \ ,\theta}
\end{eqnarray}
\noindent and
\begin{eqnarray}
\label{max2a}
&&\sin\theta\left(r^2 E^{\hat r} \right)_{,r}+
    {Hr}\left(\sin\theta E^{\hat \theta}\right)_{,\theta}
    + Hr E^{\hat \phi}_{\;,\phi}
     =  {4\pi H}r^2\sin\theta J^{\hat t}\ ,
\\
\label{max2b}
&& A\left[\left(\sin\theta  B^{\hat \phi} \right)_{,\theta}
    - B^{\hat\theta}_{\ ,\phi}\right]
     = \left({r\sin\theta}\right)
    \frac{\partial E^{\hat r}}{\partial t}
    +{4\pi}Ar\sin\theta J^{\hat r} \ ,
\\
\label{max2c}
&& AH B^{\hat r}_{\ ,\phi} - \sin\theta \left(r \ A
    B^{\hat \phi} \right)_{,r}
     =  \left({Hr\sin\theta}\right)
    \frac{\partial E^{\hat\theta}}{\partial t}
    +{4\pi AH}r\sin\theta J^{\hat\theta} \ ,
\\
\label{max2d}
&& \left(Ar B^{\hat \theta} \right)_{,r} - AH
    B^{\hat r}_{\ ,\theta}
     = \left({Hr}\right)
    \frac{\partial E^{\hat\phi}}{\partial t}
    +{4\pi AH}rJ^{\hat\phi} \ .
\end{eqnarray}

\section{Stationary Solutions to Maxwell Equations for
Magnetized Highly Conducting Spherical Star in Braneworld}
\label{ss}

    We will look for stationary
solutions of the Maxwell equation, i.e. for solutions in which we
assume that the magnetic moment of the magnetic star does not vary
in time as a result of the infinite conductivity of the stellar
medium. Because of discontinuities in the fields across the
surface of the sphere we will refer to as {\sl interior solutions}
those solutions valid within the radial range $R_{_{IN}} \le r \le
R$, and to as {\sl exterior solutions} those valid in the range $R
< r \le \infty$.  Limiting the solution to an inner radius
$R_{_{IN}}$ removes the problem of suitable boundary conditions
for $r\rightarrow 0$, and reflects the basic ignorance of the
properties of magnetic fields in the interior superconducting
regions of compact relativistic objects as neutron stars.

Assuming the magnetic field to be dipolar we look for separable
solutions of Maxwell equations (\ref{max1a})--(\ref{max1d}) and
(\ref{max2a})--(\ref{max2d}) in the form
\begin{eqnarray}
\label{ansatz_1}
&& B^{\hat r}(r,\theta) = F(r)\cos\theta \ ,
\\\nonumber\\
\label{ansatz_2}
&& B^{\hat \theta}(r,\theta) = G(r)\sin\theta \ ,
\\\nonumber\\
\label{ansatz_3}
&& B^{\hat \phi}(r,\theta) = 0 \ ,
\end{eqnarray}
\noindent where the unknown radial functions $F(r)$ and $G(r)$
will account for the relativistic corrections due to a
gravitational mass
 and the brane tension both. Magnetic field depends only from
$r$ and $\theta$ coordinates due to axial symmetry and
stationarity. The dipolar approximation is simple one for interior
field, however it is consistent with the requirement that the
field configuration should match at the boundary with external
dipolar field. Since the interior magnetic field has the dipolar
configuration, the continuity of normal and tangential components
of magnetic field at the stellar surface has to be required.

Since we are treating the interior of the star as a perfect
conductor and the exterior of the star as vacuum in the
braneworld, we can impose $J^{\hat r} = J^{\hat \theta} = J^{\hat
\phi} = 0$ in Maxwell equations~(\ref{max1a}),
(\ref{max2b})--(\ref{max2d}) and obtain Maxwell equations for the
radial part of the magnetic field as
\begin{eqnarray}
\label{ir_1}
\left(r^2 F\right)_{, r} + 2 Hr G = 0 \ , &&
\\\nonumber\\
\label{ir_2}
\left(r AG\right)_{, r} +  AH F = 0\ . &&
\end{eqnarray}

\subsection{Interior Analytical Solution}
\label{internan}

The system of equations (\ref{ir_1})--(\ref{ir_2}) can be solved
for a magnetic field which is consistent with the star's structure
and corresponds to a magnetic configuration of some astrophysical
interest.  (This is what done in the general relativistic
treatment, for instance, by Gupta et al.~\cite{getal98} in the
case of an internal dipolar magnetic field).

    Alternatively, one might specify a magnetic field
configuration and look for a compatible equation of state
for the stellar structure. In this case, the simplest possible
solution to the system (\ref{ir_1})--(\ref{ir_2}) is one
in which the magnetic field is constant throughout the
region of the star of interest and is therefore the
general relativistic analogue of it. In this case, then
\begin{equation}
\label{umf}
F=\frac{C_1}{R^3} \mu \ ,
    \hskip 3.0cm
G=-\frac{H^{-1}C_1}{R^3} \mu =-H^{-1}F\ ,
\end{equation}
where $C_1$ is an arbitrary constant whose value can be determined
after imposing the continuity across the star surface of radial
magnetic field $B^{\hat r}$, $\mu$ is the magnetic moment.

    We can now check whether the solution (\ref{umf})
is physically possible. Using (\ref{umf}) in the
equation~(\ref{ir_2}) requires that the
metric functions satisfy the condition
\begin{equation}
\label{condition}
\left(r AH^{-1} \right)_{, r} -
    AH  = 0 \ .
\end{equation}

The Einstein equation (\ref{eins1}) for the spherical star in the
braneworld gives
\begin{equation}
\label{second}
 H^2(r)=\left[1-\frac{2m(r)}{r}\right]^{-1}\ ,
\end{equation}
where the mass function is
\begin{equation}
m(r)=4\pi\int_0^r\rho^{eff}(r)r^2dr\ ,
\quad r\le R\ ,
\end{equation}
and the effective total energy density is
\begin{equation}
\rho^{eff}=\rho\left(1+\frac{\rho}{2\lambda}\right)+
\frac{6}{\kappa^4\lambda}U
\end{equation}
 and $\kappa=8\pi $.

Due to the equation (\ref{second}) the field equation
(\ref{eins1}) implies
\begin{equation}
\label{fieldeqn}
\frac{dm(r)}{dr}=4 \pi r^2\rho^{eff} \ .
\end{equation}
Then the equation (\ref{eins2}) according to (\ref{second}) can be
cast in the form
\begin{equation}
\label{frac}
\frac{A'}{A}=\frac{m(r)+4 \pi
\left(p^{eff}+\frac{4}{k^4\lambda}P\right)r^3}{r(r-2M)} \ .
\end{equation}
Inserting equation (\ref{frac}) into equation  (\ref{eins3}) one
can use it to eliminate $A'/A$ and get the modified equation of
Tolman-Oppenheimer-Volkoff for the hydrostatic
equilibrium~\cite{t39}--\cite{g96}
\begin{equation}
\label{tov1}
\frac{dp}{dr}=-\frac{(\rho+p)\left[m(r)+4 \pi\left(
p^{eff}+\frac{4}{k^4\lambda}P\right)r^3\right]}{r(r-2M)} \ .
\end{equation}

On the other hand after simple algebra one can obtain from the
Maxwell equation (\ref{condition}) using the derived expressions
(\ref{second}) and (\ref{fieldeqn}) the following expression for
hydrostatic equilibrium
\begin{equation}
\label{tov2}
\frac{dp}{dr}=-\frac{(\rho+p)\left(m(r)+4 \pi
\rho^{eff}r^3\right)}{r(r-2M)} \ ,
\end{equation}
modified by brane tension.

Comparison of the equations (\ref{tov1}) and (\ref{tov2}) gives
the following equation of state
\begin{equation}
\label{eos} \rho^{eff}=p^{eff}+\frac{4}{k^4\lambda}P
\end{equation}
for the star with the constant magnetic density, i.e. magnetic
field (\ref{umf}). If we assume that $P=0$ we will get unrealistic
exotic equation of state $\rho^{eff}=p^{eff}$ for stiff matter.

\subsection{Exterior Numerical Solution}
\label{srst_es}

    The exterior solution for the magnetic field is
simplified by the knowledge of explicit analytic expressions for
the metric functions $A$ and $H$. In particular, after defining $N
\equiv A =H^{-1}= (1 - 2M/r+Q/r^2)^{1/2}$, the system
(\ref{ir_1})--(\ref{ir_2}) can be written as a single,
second-order ordinary differential equation for the unknown
function $F$
\begin{equation}
\label{leg_eq_sim}
\frac{d}{dr}\left[\left(1-\frac{2M}{r}+\frac{Q}{r^2}\right)
    \frac{d}{dr}\left(r^2 F\right)\right] - 2F = 0 \ .
\end{equation}

The solution of equation~(\ref{leg_eq_sim}) exists when the
parameter $Q=0$~\cite{go64}--\cite{zr}. The analytical general
relativistic solution of a dipolar magnetic field in vacuum
expressed through the Legendre functions of the second
kind~\cite{aw01} shows that the magnetic field is amplified by a
factor
\begin{equation}
\label{f_of_r} \frac{F_{GR}(r)}{F_{Newt}(r)} = - \frac{3R^3}{8M^3}
    \left[\ln N^2 + \frac{2M}{r}\left(1 +  \frac{M}{r}
    \right) \right]
\end{equation}
compared to the flat space-time one. Here $F_{Newt}=2\mu/r^3$ is
the value of magnetic field at the pole in the Newtonian limit.

\begin{figure}
[htbp]
\begin{center}
\vspace{0.5cm} \hspace{-0.5cm}
 \includegraphics[width=0.5\textwidth]{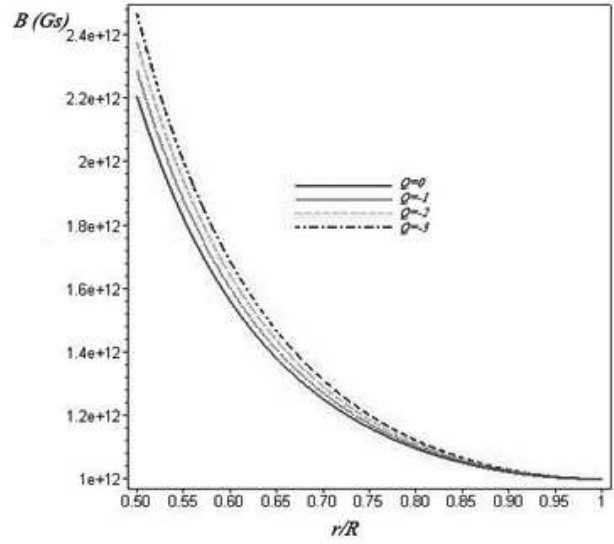}
\end{center}
\caption{A plot of magnetic field $B$ against $r/R$ in the
exterior of a relativistic magnetized compact star with typical
parameters, $M=2M_{\odot}$, $R=10~km$. The Weyl charges are taken
in the unit of $M^2$.} \label{fig:f_q}
\end{figure}

We integrate equation (\ref{leg_eq_sim}) using Runge--Kutta fifth
order method, using standard techniques befitting second--order
ODE (e.g. Press ~\cite{pre89}) in the program MAPLE 10. The ODE
equation is solved as an initial value problem. For initial values
we have chosen $B(r)=0$, $B'(r)=-\epsilon$ at $r=\infty$ taking
into account the value of magnetic star at the surface of star in
the Newtonian limit, where $\epsilon$ is small positive number. In
the limit of $r\rightarrow\infty$ the solution is taken to be
Newtonian, since $Q/r^2$ and $M/r$ are negligibly small and do not
give any contribution to the magnetic field. With such a
prescription, the equation is integrated inwards up to the surface
of the relativistic star exceeding the apparent horizon defined by
$r_+/M = 1 + \sqrt{1 - Q}$. We first reproduce the analytical form
of $F_{GR}/F_{Newt}$ (equation (\ref{f_of_r})) to better than 1
part in $10^5$ by setting $Q$ to zero. For the models in the
present study we choose the following parameters $R=10~km$ and the
polar surface field strength $B(\overline{r}=r/R=1)=10^{12}G$.
Following this, we perform the integrations for various values of
$Q$ as shown in Fig. (\ref{fig:f_q}).

    The braneworld enhancement of the exterior magnetic field at the surface
of the relativistic star is given in table~\ref{table1} which
varies between $2$ and $3$ depending on intensity of Weyl charge
selected.

\begin{table}
\caption{\label{table1}The amplification of exterior magnetic
field at the boundary surface of a typical relativistic compact
star as a result of brane tension. The Newtonian value for the
magnetic field at the pole is taken as $B_0=10^{12}~G$. The mass
of star is $M=2M_{\odot}$ and the radius is $R=10~km$. The values
for Weyl charge $Q$ are taken as usual in unit of $M^2$. }
\begin{ruledtabular}
\begin{tabular}{lcccccr}
 ${\bf Q}$ & -0.5 &  -1.0 & -1.5 & -2.0 & -2.5 & -3.0 \\
 ${\bf B,~(10^{12}~G)}$ & 1.94 & 2.05 & 2.17 & 2.32 & 2.51 & 2.77\\
\end{tabular}
\end{ruledtabular}
\end{table}

\begin{figure}
[htbp]
\begin{center}
\vspace{0.5cm} \hspace{-0.5cm}
\includegraphics[width=0.5\textwidth]{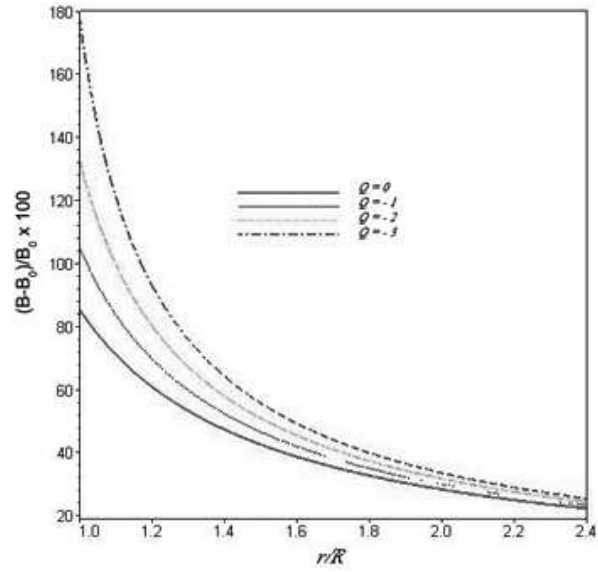}
\end{center}
\caption{The percentage increase in the magnetic field strength as
a function of radial distance $r/R$ in the exterior of
relativistic magnetized star depending on Weyl charge $Q$.}
\label{fig:fqbr2}
\end{figure}

As can be seen from the presented graphs in figure~\ref{fig:fqbr2}
the magnetic field strength outside the star will be enhanced by
braneworld parameter up to approximately $170$ percent compared
with that in the flat space-time limit depending on the value of
parameter $Q$ selected.

\subsection{Interior numerical solution}
\label{internum}

The assumption of constant density of matter is approximately
realistic because for non-exotic equations of state~\cite{bps71}
the change in density is only about one order of magnitude within
three quarters of the neutron star volume.  For uniform density
star $(\rho=const)$ one can select that $U=0$ and $P=0$. Then
according to the Einstein equations~(\ref{eins1})--(\ref{eins4})
and Maxwell equations~(\ref{ir_1})--(\ref{ir_2}) the equation for
unknown radial function will take a form
\begin{equation}
\label{eqradfun}
\frac{d}{dr}\left[\frac{\Delta(r)}{\rho+
p(r)}\frac{d}{dr}\left(r^2F\right)\right]-\frac{2F}{\left[\rho+p(r)\right]\Delta(r)}
= 0 \ ,
\end{equation}
where
\begin{eqnarray}
&& \Delta(r)\equiv
\left[1-\frac{2M}{r}\left(\frac{r}{R}\right)^3\left[1-\frac{Q}{6MR}\right]\right]^{1/2}
\ , \\ \ && p(r)=\frac{\left[\Delta(r) -
\Delta(R)\right]\left(1-Q/3MR\right)\rho}{\left[3\Delta(R)-\Delta(r)\right]-
\left[3\Delta(R)-2\Delta(r)\right]Q/3MR} \ , \\ \ && m(r) =
M\left[1 - \frac{Q}{6MR}\right]\left(\frac{r}{R}\right)^3 \ .
\end{eqnarray}

For finding the interior numerical solution we have also used the
same technique as for the exterior case, i.e the ODE is considered
as an initial value problem. For initial value of $B
(\overline{r}=1)$ taken the value obtained from the exterior
solution due to assumed continuity of magnetic field through the
surface of the star.
The Maxwell equation is integrated up to $\overline{r}=0.5$ where
the existence of the dipolar magnetic field is expected.

\begin{figure}
[htbp]
\begin{center}
\vspace{0.5cm} \hspace{-0.5cm}
 \includegraphics[width=0.5\textwidth]{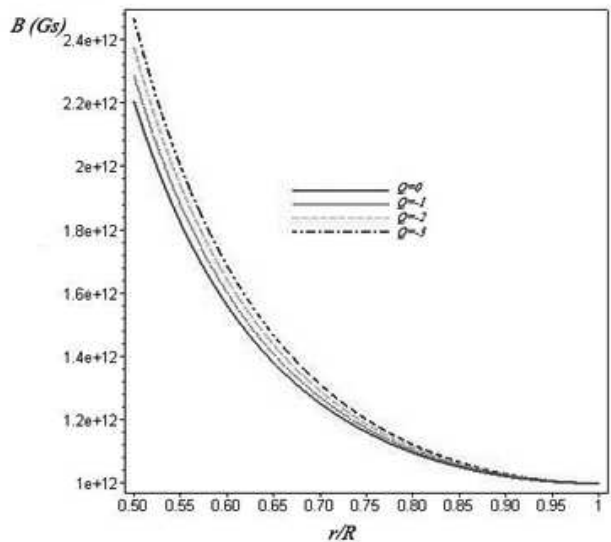}
\end{center}
\caption{A plot of magnetic field $B$ against $r/R$ for the
interior region of typical compact relativistic star with
parameters $M=2~M_{\odot}$, $R=10 ~km$ and $\rho = 1.0\times
10^{14}$~g~cm$^{-3}$ for various values of $Q$. The brown solid
curve is for $Q=0$, the red dotted curve for $Q=-1$, the green
dashed curve for $Q=-2$ and the blue dot--dashed curve for $Q=-3$
which are taken in the unit of $M^2$.} \label{fig:srho}
\end{figure}

In the figure~\ref{fig:srho} we present the magnetic field
strength of the interior of a star as a function of the radial
distance from $\overline{r}=0.5$ (which is normalized on the
radius of star $R$) up to the stellar surface $\overline{r}=1$.
One can see from the presented graphs the field strength interior
of the star will be enhanced by braneworld parameter and at the
half radius of the star is increased by up to approximately $20$
percent compared with that for the general-relativistic
Schwarzschild star depending on the value of parameter $Q$
selected.

\section{Astrophysical Consequences}
\label{application}

Assume that the oblique magnetized braneworld star is rotating and
$\chi$ is the inclination angle between axis of rotation and
magnetic momentum and observed as pulsar through magnetic dipole
radiation. Then the luminosity of the relativistic star in the
case of a purely dipolar radiation, and the power radiated in the
form of dipolar electromagnetic radiation, is given by~\cite{ra04}
\begin{equation}
\label{dipole_energy_loss} L_{em} = \frac{\Omega^4_{_R} R^6
{\widetilde B}^2_0}{6 c^3}\sin^2\chi
    \ ,
\end{equation}
where subscript $0$ denotes the value at $r=R$.

When compared with the equivalent Newtonian expression for the
rate of electromagnetic energy loss through dipolar radiation
\cite{ll87}
\begin{equation}
\label{dipole_energy_loss_newt} (L_{em})_{\rm Newt} =
    \frac{\Omega^4 R^6 B_0^2}{6 c^3}\sin^2\chi
    \ ,
\end{equation}
it is easy to realize that the general relativistic braneworld
corrections emerging in expression (\ref{dipole_energy_loss}) are
partly due to the magnetic field amplification ${\widetilde B}_0 =
F_{_R} B_0$ at the stellar surface and partly to the increase in
the effective rotational angular velocity produced by the
gravitational redshift as $\Omega = \Omega_{_R} A_{_R}$.

\begin{figure}
[htbp]
\begin{center}
\vspace{0.5cm} \hspace{-0.5cm}
\includegraphics[width=0.5\textwidth]{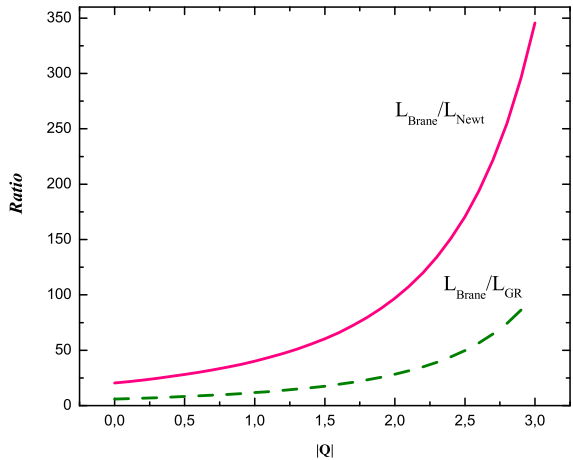}
\end{center}
\caption{General relativistic amplification of the energy loss
through dipolar electromagnetic radiation as a function of
absolute value of Weyl charge. The solid line refers to the ratio
of energy loss in the braneworld with energy loss in the flat
spacetime, while dashed line is with energy loss in the
Schwarzschild curvature background. It is shown with a dashed line
for comparison. Weyl charge $|Q|$ in abscissa is taken in the unit
of $M^2$.} \label{fig1}
\end{figure}

The presence of a braneworld tension has the effect of enhancing
the rate of energy loss through dipolar electromagnetic radiation
by an amount which can be easily estimated to be
\begin{equation}
\label{dipole_energy_loss_cf} \frac{L_{em}}{(L_{em})_{_{\rm
Newt}}}=
        \left(\frac{F_{_R}}{A^2_{_R}}\right)^2\ ,
\end{equation}
and whose dependence is shown in figure~\ref{fig1} with a solid
line. Considering expression (\ref{dipole_energy_loss}) in the
selected range for the brane tension, it is straightforward to
realize that the Newtonian expression
(\ref{dipole_energy_loss_newt}) underestimates the electromagnetic
radiation losses of a factor which may reach few hundreds
depending on the value of the brane tension parameter.

\begin{figure}
[htbp]
\begin{center}
\vspace{0.5cm} \hspace{-0.5cm}
\includegraphics[width=0.5\textwidth]{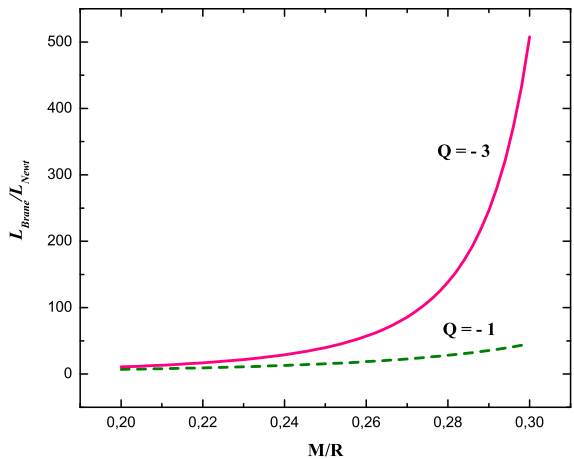}
\end{center}
\caption{General relativistic amplification of the energy loss
through dipolar electromagnetic radiation as a function of stellar
compactness $M/R$ for different Weyl charges.} \label{fig2}
\end{figure}

    In the figure~\ref{fig2} we present enhancement of
the electromagnetic energy loss by brane tension depending on
compactness of the relativistic star and prove that the braneworld
effects are more important for ultracompact stars.

    Expression (\ref{dipole_energy_loss})  could be
used to investigate the rotational evolution of magnetized neutron
stars with predominant dipolar magnetic field anchored in the
crust which converting its rotational energy into electromagnetic
radiation. A detailed  investigation of general relativistic
effects for Schwarzschild stars has already been performed by Page
et al.~\cite{pgz00}, who have paid special attention to the
general relativistic corrections that need to be included for a
correct modeling of the thermal evolution but also of the magnetic
and rotational evolution. It should be remarked, however, that in
their treatment Page et al.~\cite{pgz00} have taken into account
the general relativistic amplification of magnetic field  due to
the curved background spacetime, but did not include the
corrections due to the gravitational redshift. As a result, the
general relativistic electromagnetic luminosity estimated by Page
et al.~\cite{pgz00} is smaller than the one computed in our
paper~\cite{ra04} where all general relativistic effects are taken
into account.

\section{Conclusion}
\label{conclusion}

We have studied the stationary magnetic field of isolated
relativistic compact star in the braneworld assuming that their
magnetic fields are confined to the stellar crust of ideal perfect
matter and have been working with the braneworld effects on the
stellar magnetic field, accompanied by proper boundary conditions.
In other words we generalized the general relativistic approach in
the sense that we took into account the effect of additional
braneworld tension on electromagnetic fields.

First we have solved interior Maxwell equations analytically and
found exact solution for interior magnetic field inside the
stellar stiff matter with unrealistic equation of state.

Then we have found numerical calculations which take into account
the effect of brane tension on the structure of magnetic field
outside the star and configuration of interior magnetic field for
the stellar matter with constant density. Comparing the behaviour
of the magnetic field when brane tension effects are included with
the case when they are neglected, one can see enhancement of
magnetic field especially near the surface of the relativistic
star for external field and in the inner boundary taken for the
interior field. This effect is stronger the more bigger the
braneworld parameter $Q$ is.

Numerically calculated magnetic field structure for the external
magnetic field, without brane tension effects incorporated,
totally match the known analytical solution for magnetic field
when 'branetension charge' $Q$ is equal to zero.

The numerical calculations made confirm that there are two effects
of brane tension on magneto-dipolar emission. One is due to
amplification of surface magnetic field by brane tension. Other is
due to the presence of function $Q/r^2$ in the red-shift factor
$\sqrt{1-2M/r+Q/r^2}$ defined at the stellar surface in the
expression for power of magnetodipolar radiation. We have found
that the effect of branetension on magnetic fields of compact
stars can be very important, and the expression for magnitodipolar
luminosity of rotating braneworld magnetized star gives
enhancement up to two orders.

While the previous papers extensively studied the influence of
braneworld effects into the different astrophysical processes and
cosmological problems, we conclude here the incorporation of the
brane tension effect into magnetic field structure of the
relativistic stars, could give an additional key for astrophysical
evidence of the parameter $Q$.

\section*{Acknowledgments}

The authors thank Naresh Dadhich for setting the problem and
useful comments, M Sami for involving us to the braneworld subject
and discussions, and Arun Thampan for his help in performing
numerical calculations. FJF acknowledges the fellowship from the
ICTP-TRIL program. BJA is grateful to TWAS for the travel support
and to the IUCAA for warm hospitality during his stay in Pune.
This research is also supported in part by the UzFFR (project
01-06) and projects F.2.1.09, F2.2.06 and A13-226 of the UzCST.
BJA acknowledges the partial financial support from NATO through
the reintegration grant EAP.RIG.981259.


%

\end{document}